\documentclass[final,1p,times]{elsarticle}
 \usepackage{graphics}
\usepackage{amssymb}
\journal{Physica D}

\begin{document}

\begin{frontmatter}



\title{Surface solitons in trilete lattices}

\author[nis]{M.˜Stojanovi\'c}
\author[nis]{A.˜Maluckov\corref{cor1}}
\ead{sandra@pmf.ni.ac.rs}\cortext[cor1]{Corresponding author}
\author[vinca]{Lj.˜Had\v zievski}
\author[izr]{B.A.˜Malomed}
\address[nis]{Faculty of Sciences and Mathematics, University of Ni\v s, P.O.B. 224,
18000 Ni\v s, Serbia}
\address[vinca]{Vin\v ca Institute of Nuclear Sciences, University of Belgrade, P.O.
Box 522,11001 Belgrade, Serbia}
\address[izr]{Department of Physical Electronics, School of Electrical Engineering,
Faculty of Engineering, Tel Aviv University, Tel Aviv 69978,
Israel}

\begin{abstract}
Fundamental solitons pinned to the interface between three
semi-infinite one-dimensional nonlinear dynamical chains, coupled
at a single site, are investigated. The light propagation in the
respective system with the self-attractive on-site cubic
nonlinearity, which can be implemented as an array of nonlinear
optical waveguides, is modeled by the system of three discrete
nonlinear Schr\"{o}dinger equations. The formation, stability and
dynamics of symmetric and asymmetric fundamental solitons centered
at the interface are investigated analytically by means of the
variational approximation (VA) and in a numerical form. The VA
predicts that two asymmetric and two antisymmetric branches exist
in the entire parameter space, while four asymmetric modes and the
symmetric one can be found below some critical value of the
inter-lattice coupling parameter -- actually, past the
symmetry-breaking bifurcation. At this bifurcation point, the
symmetric branch is destabilized and two new asymmetric soliton
branches appear, one stable and the other unstable. In this area,
the antisymmetric branch changes its character, getting stabilized
against oscillatory perturbations. In direct simulations, unstable
symmetric modes radiate a part of their power, staying trapped
around the interface. Highly unstable asymmetric modes transform
into localized breathers traveling from the interface region
across the lattice without significant power loss.
\end{abstract}


\end{frontmatter}

% \linenumbers

\section{Introduction}
\label{} Surface modes, which are a special type of waves
localized at interfaces between different media, were first
predicted as localized Tamm electronic states at the edge of a
truncated periodic potential \cite{rad1}. In optics, it was
predicted theoretically and confirmed experimentally that the
nonlinear self-trapping of light near the edge of a waveguide
array with the self-focusing nonlinearity can lead to the
formation of discrete surface solitons \cite{rad2,rad3}. Various
settings for the creation of surface solitons were also proposed
for Bose-Einstein condensates (BECs) \cite{BEC}. The general
framework for the description of such localized patterns is
provided by the discrete nonlinear-Schr\"{o}dinger (DNLS)
equations, with appropriate boundary conditions \cite{book}.

The solitary surface modes exist above a certain threshold power
and, in a certain domain of the parameter space, different surface
modes can exist simultaneously. The surface modes may be
understood as discrete optical solitons localized near the surface
\cite{rad4}-\cite{rad5}, or as lattice solitons pinned by defects
\cite{rad5a}-\cite{rady}. Surface solitons supported by truncated
superlattices were investigated too \cite{He-super}. Moreover, the
light localization in self-defocusing nonlinear media in the form
of surface gap solitons has been predicted and observed in Refs.
\cite{rad7,rad8,rad8a,rad6}, and the concept of multi-gap surface
solitons, i.e., mutually trapped surface modes with components
associated with different spectral gaps, was put forward
\cite{rad6,rad6a} (multi-gap, alias "inter-gap", or "semi-gap",
solitons are also known in uniform lattice media \cite{inter}). A
short review of surface solitons in discrete systems was given in
Ref. \cite{He}.

The studies of surface modes have shown that nonlinear discrete
photonic and matter-wave systems support spatially localized
states with sundry symmetries (which can be controlled by the
insertion of suitable defects into the system) \cite{rad5d,bec}.
Related to this is the possibility of the \textit{spontaneous
symmetry breaking} (SSB) in symmetric dual-core systems, with a
linear coupling between the two parallel cores. In fact, the SSB
bifurcation, which destabilizes symmetric states and gives rise to
asymmetric ones, was originally predicted in terms of the
self-trapping in discrete systems \cite{13}. In the physically
important model of dual-core
nonlinear optical fibers, the SSB instability was discovered in Ref. \cite%
{14}, and the respective bifurcations for continuous-wave states
were studied in detail in Ref. \cite{15}, for various types of the
intra-core nonlinearities. Further, the SSB was studied for
solitons (rather than continuous waves) in the model of the
dual-core fiber with the cubic (Kerr)
nonlinearity \cite{18,17}, for gap solitons in the models of dual-core \cite%
{19} and tri-core \cite{20} fiber Bragg gratings, and for
matter-wave solitons in the BEC loaded into a dual-core potential
trap, that may be combined with a longitudinal optical lattice
\cite{26a}. In addition, the
SSB was also analyzed in models describing optical media with quadratic \cite%
{25} and cubic-quintic \cite{26} nonlinearities.

The variational approximation (VA) \cite{Progress} makes it
possible to study the SSB in dual-core systems in an analytical
form \cite{18,VA}. It is relevant to mention that the VA may allow
one not only to describe
fundamental localized modes, but also correctly predict their stability \cite%
{26b}.

Continuing the investigation of the surface fundamental modes in
coupled one-dimensional (1D) lattice system \cite{VA,symnash}, in
this work we study soliton complexes formed at the interface of
three identical semi-infinite lattices which form a trilete
configuration, see Fig. 1 below. Such lattices can be written in
bulk silica by means of femtosecond optical pulses, which is the
technique which has made the creation of other types of surface
solitons possible \cite{Jena}. A major motivation for studying
such trilete systems is that they provide a fundamental setting
for the study of the SSB
which is different from the well-studied dual-core configurations \cite%
{20,Skryabin}.

The present system is modeled by three DNLS equations coupled at a
single lattice site. In the framework of this system, we consider
complexes built of three fundamental surface solitons, each
carried by one of the three chains. Actually, these fundamental
solitons are realizations of Tamm states in the present setting.
We show that they form families of symmetric and asymmetric
complexes, which exhibit the power-threshold behavior. Their
stability and propagation dynamics are examined in detail, using
the VA and numerical calculations. The existence regions for all
families of the soliton complexes are produced by means of both
approaches, the analytical results being quite close to the
numerical ones. Stability properties, predicted by dint of the
numerically implemented linear-stability analysis, are verified by
direct numerical simulations of the soliton evolution.

The rest of the paper is structured as follows. The model is
formulated in Section 2, which is followed by the consideration of
the VA in Section 3. Numerical results are collected in Section 4,
and the paper is concluded by Section 5.

\section{The model}
\begin{figure}[h]
\center\includegraphics [width=4cm]{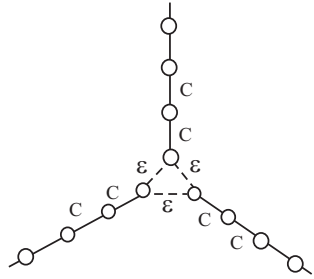}
\caption{Three coupled chains, linked by the modified linear coupling $%
\protect\varepsilon $, form the trilete lattice.} \label{fig1}
\end{figure}

As said above, the trilete lattice configuration is formed by
three identical semi-infinite sublattices linked at one site. The
inter-site coupling constant is scaled to be $C\equiv 1$ inside
the lattices, while a different constant, $\varepsilon $, accounts
for the linkage between the lattices, having the same sign as $C$,
see Fig. \ref{fig1}. Thus, the triple lattice is modeled by the
following system of three coupled DNLS equations,
\begin{eqnarray}
i\frac{d\phi _{n}}{dz}+\left( \phi _{n+1}+\phi _{n-1}\right)
-\delta _{n,0}\phi _{n-1}+\varepsilon \delta _{n,0}(\psi
_{n}+\theta
_{n})+\left\vert \phi _{n}\right\vert ^{2}\phi _{n} &=&0,  \nonumber \\
i\frac{d\psi _{n}}{dz}+\left( \psi _{n+1}+\psi _{n-1}\right)
-\delta _{n,0}\psi _{n-1}+\varepsilon \delta _{n,0}(\phi
_{n}+\theta
_{n})+\left\vert \psi _{n}\right\vert ^{2}\psi _{n} &=&0,  \label{system} \\
i\frac{d\theta _{n}}{dz}+\left( \theta _{n+1}+\theta _{n-1}\right)
-\delta _{n,0}\theta _{n-1}+\varepsilon \delta _{n,0}(\phi
_{n}+\psi _{n})+\left\vert \theta _{n}\right\vert ^{2}\theta _{n}
&=&0,  \nonumber
\end{eqnarray}%
where $z$ is the propagation distance (assuming that the chains
represent three semi-infinite arrays of optical waveguides),
$n=0,1,...,N$ is the discrete coordinate in the chain ($N$ is the
total number of sites in each lattice that was used in actual
numerical calculations), $\delta _{n,0}$ is the Kronecker's
symbol, and rescaling is used to make the on-site self-attraction
coefficient equal to $1$.

Soliton solutions to Eqs. (\ref{system}) are looked for in the usual form,%
\begin{equation}
\phi _{n}=u_{n}\exp \left( i\mu z\right) ,~\psi _{n}=v_{n}\exp
\left( i\mu z\right) ,~\theta _{n}=w_{n}\exp \left( i\mu z\right)
,  \nonumber
\end{equation}%
where $u_{n}$, $v_{n}$ and $w_{n}$ are real discrete functions,
and $\mu $ is the propagation constant. The corresponding
stationary equations,
\begin{eqnarray}
-\mu u_{n}+u_{n+1}+u_{n-1}-\delta _{n,0}u_{n-1}+\varepsilon
(v_{n}+w_{n})\delta _{n,0}+u_{n}^{3} &=&0,  \nonumber \\
-\mu v_{n}+v_{n+1}+v_{n-1}-\delta _{n,0}v_{n-1}+\varepsilon
(u_{n}+w_{n})\delta _{n,0}+v_{n}^{3} &=&0,  \label{eq3} \\
-\mu w_{n}+w_{n+1}+w_{n-1}-\delta _{n,0}w_{n-1}+\varepsilon
(u_{n}+v_{n})\delta _{n,0}+w_{n}^{3} &=&0,  \nonumber
\end{eqnarray}%
can be derived from the Lagrangian,
\begin{equation}
L=L_{u}+L_{v}+L_{w}+2\varepsilon \left(
u_{0}v_{0}+v_{0}w_{0}+w_{0}u_{0}\right) ,  \label{eq4}
\end{equation}%
where the intrinsic Lagrangians of the three semi-infinite chains
are, respectively,
\begin{eqnarray}
L_{u} &\equiv &\sum_{n=0}^{\infty }\left( -\mu u_{n}^{2}+\frac{1}{2}%
u_{n}^{4}+2u_{n}u_{n+1}\right) ,  \nonumber \\
L_{v} &\equiv &\sum_{n=0}^{\infty }\left( -\mu v_{n}^{2}+\frac{1}{2}%
v_{n}^{4}+2v_{n}v_{n+1}\right) ,  \nonumber \\
L_{w} &\equiv &\sum_{n=0}^{\infty }\left( -\mu w_{n}^{2}+\frac{1}{2}%
w_{n}^{4}+2w_{n}w_{n+1}\right) ,  \nonumber
\end{eqnarray}%
and the last term in Eq. (\ref{eq4}) accounts for the coupling
between them.

\section{The variational approximation}

The analytical approach to the study of the soliton solutions may
be based on the VA\ (variational approximation) \cite{Progress},
which was adapted to
discrete systems in several earlier works \cite{VA,26b,symnash,discreteVA,mi}%
. Following this method, we adopt the following \textit{ansatz}:
\begin{equation}
\left\{ u_{n},v_{n},w_{n}\right\} =\left\{ A,B,C\right\} \exp
{(-an)},~\ \mathrm{at~}n\geq 0,  \label{eq6}
\end{equation}%
with amplitudes $A$, $B$ and $C$ treated as variational
parameters. As concerns inverse width $a$, following Ref.
\cite{Kaup} we fix it through a solution of the linearized version
of Eqs. (\ref{eq3}) for ``tails" of the discrete solitons at
$n\rightarrow \infty $, from where it follows
\begin{equation}
s\equiv e^{-a}=\mu /2-\sqrt{\left( \mu /2\right) ^{2}-1},
\label{eq8}
\end{equation}%
hence the solitons may exist for values of the propagation
constant $\mu \geq 2$. Note that Eq. (\ref{eq8}) yields $s<1$,
which is essential for the
validity of analytical results displayed below -- see Eqs. (\ref{eq13}) and (%
\ref{eq13arr}).

The substitution of ansatz (\ref{eq6}) into Eq. (\ref{eq4}) yields
the corresponding effective Lagrangian,
\begin{equation}
L=L_{1}+L_{2}+L_{3}+2\varepsilon \left( AB+AC+BC\right) ,
\label{eq9}
\end{equation}%
\begin{eqnarray}
L_{1} &=&A^{2}\frac{-\mu +2s}{1-s^{2}}+\frac{1}{2\left(
1-s^{4}\right) }A^{4}
\nonumber \\
L_{2} &=&B^{2}\frac{-\mu +2s}{1-s^{2}}+\frac{1}{2\left(
1-s^{4}\right) }B^{4}
\nonumber \\
L_{3} &=&C^{2}\frac{-\mu +2s}{1-s^{2}}+\frac{1}{2\left( 1-s^{4}\right) }%
C^{4}.  \nonumber
\end{eqnarray}%
The Euler-Lagrange equations for amplitudes $A$, $B$ and $C$ are
derived from this Lagrangian in the form of
\begin{eqnarray}
\frac{\partial L_{1}}{\partial A}+2\varepsilon \left( B+C\right)
&=&0,
\nonumber \\
\frac{\partial L_{2}}{\partial B}+2\varepsilon \left( A+C\right)
&=&0,
\nonumber \\
\frac{\partial L_{3}}{\partial C}+2\varepsilon \left( A+B\right)
&=&0, \nonumber
\end{eqnarray}%
or, in the explicit form,
\begin{eqnarray}
-\frac{1}{s}A+\frac{1}{1-s^{4}}A^{3}+\varepsilon \left( B+C\right)
&=&0,
\nonumber \\
-\frac{1}{s}B+\frac{1}{1-s^{4}}B^{3}+\varepsilon \left( A+C\right)
&=&0,
\label{eq11} \\
-\frac{1}{s}C+\frac{1}{1-s^{4}}C^{3}+\varepsilon \left( A+B\right)
&=&0. \nonumber
\end{eqnarray}

These equations allow us to predict the existence of different
symmetric and asymmetric interface modes.

\subsection{Existence regions for the interface solitons}

\begin{figure}[h]
\center\includegraphics [width=6cm]{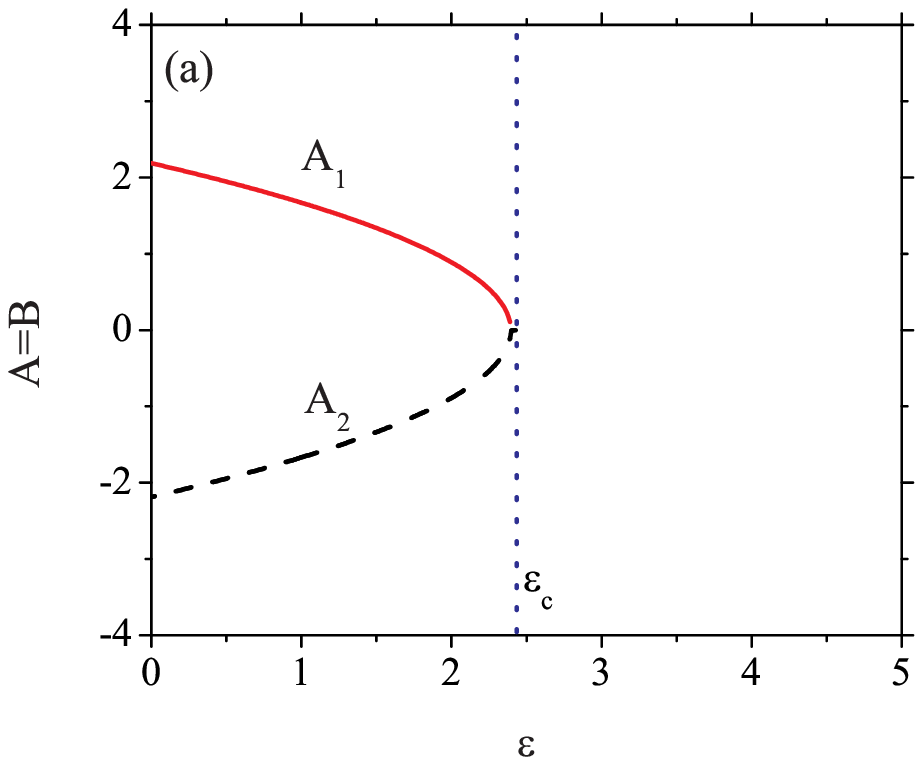}%
\includegraphics
[width=6cm]{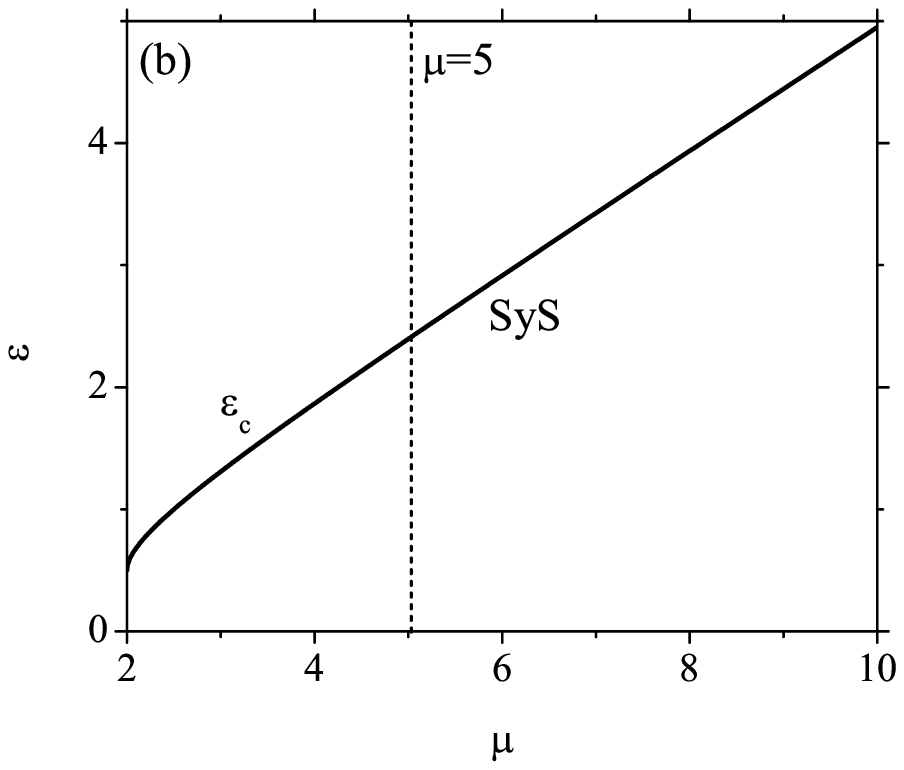} \caption{(color online) Results of the
variational approximation for symmetric solitons. (a) The
amplitude versus the inter-chain coupling
constant, $\protect\varepsilon $, for a fixed propagation constant, $\protect%
\mu =5$, i.e., $s=0.21$ (the same value is used in all examples
displayed below). In this panel, $A_{2}\equiv -A_{1}$, see Eq.
(\protect\ref{eq13}). (b) In the $(\protect\mu
,\protect\varepsilon )$ plane, the existence region
for the fundamental symmetric solitons, marked by symbol "SyS", is $\protect%
\varepsilon <\protect\varepsilon _{\mathrm{c}}=1/(2s)$. For $\protect\mu =5$%
, $\protect\varepsilon _{\mathrm{c}}\approx 2.4$.} \label{fig2}
\end{figure}

The solution for symmetric solitons, with $A=B=C$, is easily
obtained from
Eqs. (\ref{eq11}):%
\begin{equation}
A=\pm \sqrt{({1-s^{4}})(s^{-1}-2\varepsilon }).  \label{eq13}
\end{equation}%
The dependence of this amplitude on $\varepsilon $ for fixed $\mu
=5$, i.e.,
$s=\left( 5-\sqrt{21}\right) /2\approx \allowbreak 0.21$, is plotted in Fig. %
\ref{fig2}(a). It follows from Eq. (\ref{eq13}) and is shown in
Fig. \ref{fig2}(b) that the existence of the VA-predicted
symmetric solutions at a given value of $\mu $ is limited to
\begin{equation}
\varepsilon <\varepsilon _{\mathrm{c}}=(2s)^{-1}\equiv \left( \mu
-\sqrt{\mu ^{2}-4}\right) ^{-1}.  \label{c}
\end{equation}

The variational calculations have shown that the SyS complexes
with very small power are created near the $\varepsilon_c=1/(2s)$.
In this parameter region, when the nonlinear interaction is
negligible, we can actually analyze the corresponding linear
trilete lattice system. The straightforward analysis shows that
only symmetric linear surface localized complexes with arbitrary
amplitude can be formed in the linear trilete lattice system.
These linear modes exist exactly at $\varepsilon=1/(2s)$, which
coincides with the critical value of the inter-lattice coupling
constant ($\varepsilon_c$) for existence of symmetric nonlinear
modes. Therefore, we expect that the stable nonlinear SyS branch
with small power emerges from the linear symmetric localized mode.

\begin{figure}[h]
\center\includegraphics [width=11cm]{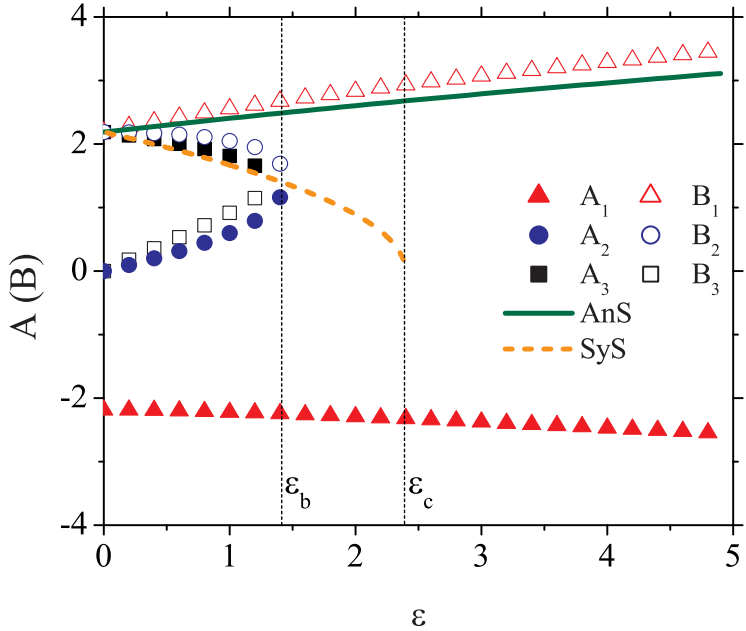} \caption{(color
online) Black, red, and blue symbols display amplitudes $A=C$
and $B$ of different branches of asymmetric solutions of the \textit{%
isosceles} type, which are predicted by the variational
approximation. Only positive branches of $B$, but not their
negative counterparts, are shown. The olive solid line denotes the
antisymmetric ("AnS")
branch, with $A=0,\,B=-C\neq 0$, which corresponds to Eq. (\protect\ref%
{eq13arr}), while the dashed orange line denotes the amplitude of
the symmetric solitons ("SyS"). Each solution has its negative
counterpart (see the text). The critical values of the coupling
constant,
corresponding to this $s$, are displayed too by vertical lines, $\protect%
\varepsilon _{\mathrm{b}}=2/(7s)$ and $\protect\varepsilon _{\mathrm{c}%
}=1/(2s)$.} \label{fig3}
\end{figure}

The critical value of $\varepsilon $ at which the branch of
asymmetric
solitons bifurcates from the symmetric family is predicted by solving Eqs. (%
\ref{eq11}) for the soliton's amplitudes with infinitesimal
differences between them: $B=A+\delta B,\,C=A+\delta C$.
Straightforward algebraic manipulations, which include the
linearization with respect to $\delta B$ and $\delta C$, yield the
value of $\varepsilon $ at the SSB bifurcation point: $\varepsilon
_{\mathrm{b}}=2/(7s)$, see Fig. \ref{fig3}. Asymmetric solutions
exist if the linear coupling is weaker than at the bifurcation
point, i.e., at $\varepsilon <\varepsilon _{\mathrm{b}}$.

An analytical solution for asymmetric modes can be easily found in
the particular case when one amplitude vanishes, $A=0$. Then two
other amplitudes are
\begin{equation}
B=-C=\pm \sqrt{(1-s^{4})(s^{-1}+\varepsilon )}.  \label{eq13arr}
\end{equation}%
This solution may be naturally called antisymmetric. Note that,
unlike symmetric solution (\ref{eq8}), it exists for all values of
$\varepsilon $. In fact, the antisymmetric solution is identical
to its counterpart found in the two-chain system in Ref.
\cite{mi}, although this does not mean that the stability
properties of the antisymmetric solutions are identical in the two
systems.

Another analytical asymmetric solution (which may be called an \textit{%
isosceles} mode) has $A=C\neq B\neq 0$. In this case, $B$ can be
eliminated,
\begin{equation}
B_{1,2}=\frac{1}{2}\left[ -A\pm
\sqrt{A^{2}-4(A^{2}-(s^{-1}+\varepsilon )(1-s^{4}))}\right] ,
\label{eq15}
\end{equation}%
while $A$ has to be found as a numerical solution of the remaining
equation,
\begin{equation}
A^{3}+\left( 1-s^{4}\right) \left( \frac{\varepsilon
}{2}-\frac{1}{s}\right)
A=\pm \frac{\varepsilon }{2}\left( 1-s^{4}\right) \sqrt{%
-3A^{2}+4(1-s^{4})(s^{-1}+\varepsilon )}.  \label{eq14}
\end{equation}%
In accordance with the above result that $\varepsilon
_{\mathrm{b}}=2/(7s)$ determines the location of the SSB
bifurcation, six different branches of solutions to Eq.
(\ref{eq14}) (three with positive and three with negative $B
$) exist in the area of $\varepsilon <2/(7s)$ of parameter space $%
(\varepsilon ,\mu )$, while only two branches (one with positive
and one with negative $B$) are found at $\varepsilon >2/(7s)$, as
shown in Fig. \ref{fig3} for $\mu =5$. In this figure, only
branches with $A=C$ and $B>0$ are shown. It is worthy to mention
that, in the region of $0<\varepsilon <\varepsilon _{\mathrm{b}}$
in the figure, all the asymmetric branches, with numbers 1, 2, and
3, coexist with the symmetric branch and the antisymmetric
one with $A=0$; further, in the region of $\varepsilon _{\mathrm{b}%
}<\varepsilon <\varepsilon _{\mathrm{c}}$, only branch $A_{1}$
coexists with the same pair, while in the region of $\varepsilon
_{\mathrm{c}}<\varepsilon $ only branches $A_{1}$ and $A=0$ exist.
It is possible to prove that
solutions of Eq. (\ref{eq11}) with all the three amplitudes different, $%
A\neq B\neq C$, do not exist, which was also confirmed
numerically.

Finally, it is relevant to mention that the asymptotic form of the
stationary solutions can be easily understood for both
$\varepsilon \rightarrow 0$ and $\varepsilon \rightarrow \infty $.
Indeed, in the limit of $\varepsilon \rightarrow 0$ the system
splits into three uncoupled semi-infinite lattices. In this case,
for fixed propagation constant $\mu $, one may have either the
usual single-component surface soliton \cite{rad2}, or the zero
solution. Accordingly, Fig. \ref{fig3} demonstrates that the
amplitudes of all modes at $\varepsilon =0$ take either a fixed
value, which actually corresponding to the single semi-infinite
lattice, or vanish.

In the limit of $\varepsilon \rightarrow \infty $, a
straightforward analysis of Eqs. (\ref{eq11}) yields the following
explicit solution, which
represents the isosceles mode in this limit:%
\[
A\approx \mp \alpha \sqrt{\varepsilon \left( 1-s^{4}\right)
},B\approx \pm \beta \sqrt{\varepsilon \left( 1-s^{4}\right) },
\]%
where $\beta \approx 1.138$ is a root of equation $\left( \beta
^{2}\right) ^{3}-2\left( \beta ^{2}\right) ^{2}+4\beta ^{2}-4=0$,
and $\alpha =\beta ^{3}/2\approx 0.737$. This expression is
similar to the\ asymptotic limit of Eq. (\ref{eq13arr}),
$B=-C\approx \sqrt{\varepsilon \left( 1-s^{4}\right) }$.

\subsection{Stability of the fundamental solitons (the
Vakhitov-Kolokolov criterion)}

The stability of the discrete solitons predicted by the VA can be
estimated by dint of the Vakhitov-Kolokolov (VK) criterion,
according to which the
necessary condition for the stability of fundamental solitons is $dP/d\mu >0$%
, where $P$ is the total power (norm) of the soliton
\cite{discreteVA,isrl}. The total power of the solutions
corresponding to ansatz (\ref{eq6}) is
\begin{equation}
P\equiv \sum_{n=0}^{\infty }(u_{n}^{2}+v_{n}^{2}+w_{n}^{2})=\frac{1}{1-s^{2}}%
({A^{2}}+B^{2}+C^{2}),  \label{eq16}
\end{equation}%
where the relation between $s$ and $\mu $ is given by Eq.
(\ref{eq8}).

\begin{figure}[h]
\center\includegraphics [width=4cm]{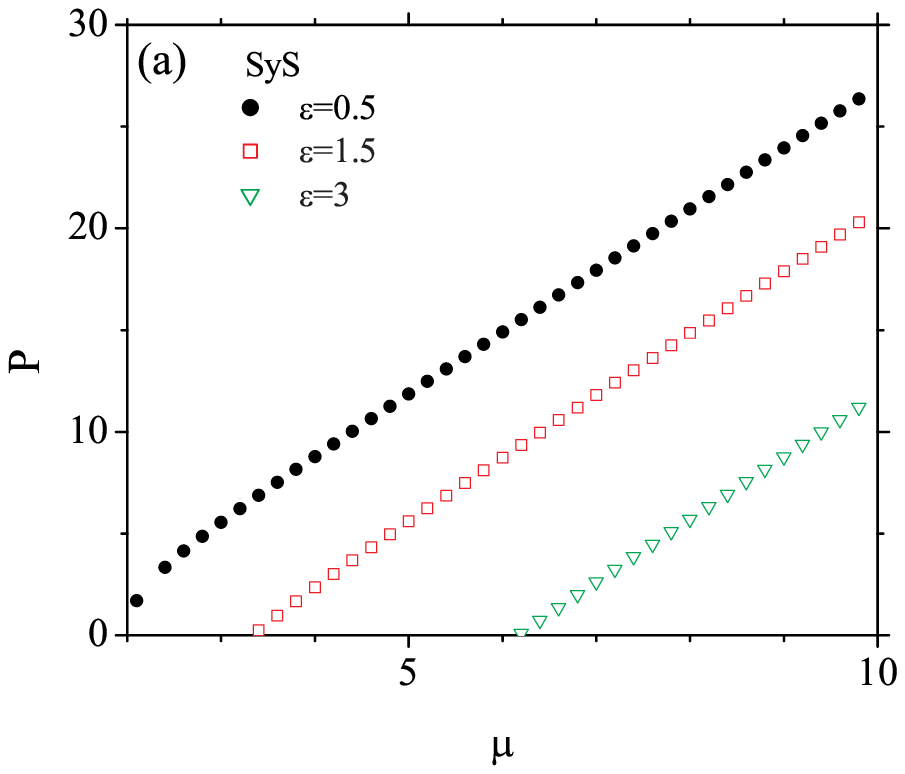}%
\includegraphics
[width=4cm]{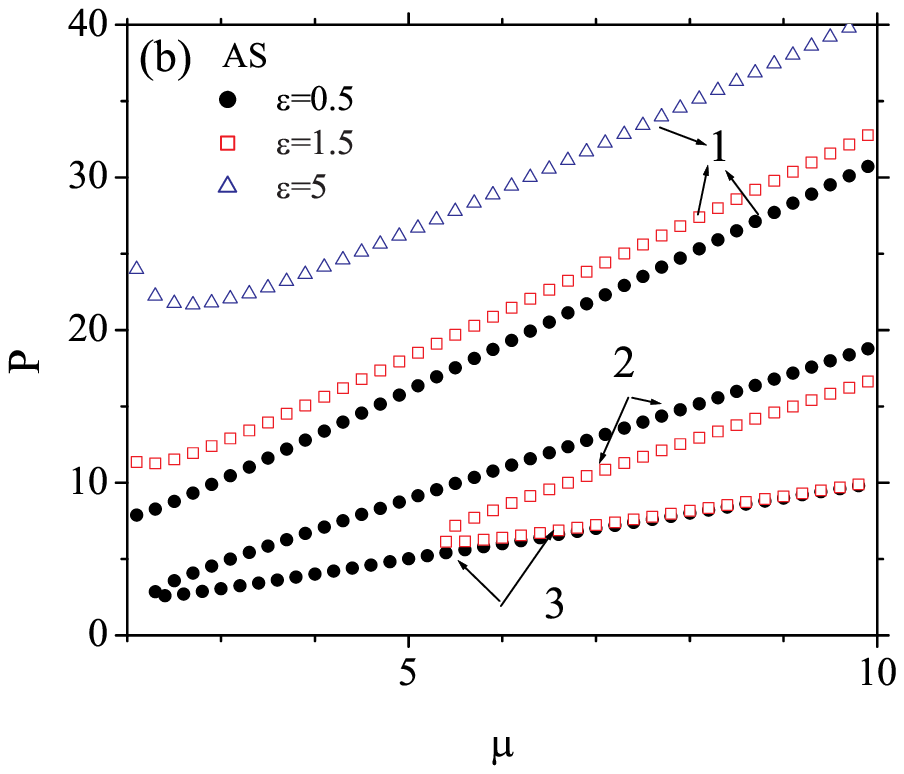}\includegraphics [width=4cm]{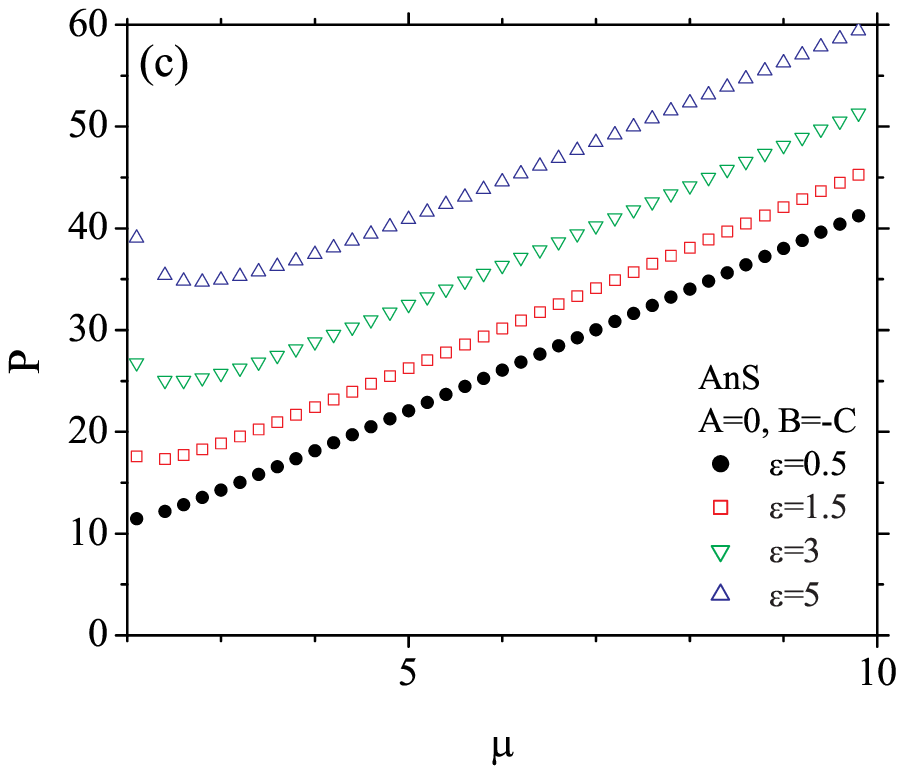}
\caption{The norm (power) $P$ versus propagation constant
$\protect\mu $ for (a) symmetric solutions ("SyS"), (b) asymmetric
(isosceles) solutions ("AS"), and (c) antisymmetric solutions,
with $A=0,\,B=-C\neq 0$ ("AnS").
The corresponding values of the inter-lattice coupling constant, $\protect%
\varepsilon $, are indicated in the panels. Different curves
labeled by indices $1,2,3$ for fixed $\protect\varepsilon $ in
panel (b) correspond to solutions with the same numbers in Fig.
\protect\ref{fig3}.} \label{fig4}  %vkfig
\end{figure}

For the symmetric solution (\ref{eq13}) with $A=B=C$, dependence
$P(\mu )$
is plotted in Fig. \ref{fig4}(a) for fixed $\varepsilon =0.5,\,1.5$ and $3$%
. The slope of the $P(\mu )$ curve is positive in the whole
existence region of the symmetric solutions, which, according to
the VK criterion, indicates the possibility for the existence of
stable solutions.

The $P(\mu )$ curve for the asymmetric solution branches is plotted in Fig. %
\ref{fig4}(b) for three different values of $\varepsilon $. Only
the asymmetric branch which corresponds to amplitudes $\left\{
A_{1},B_{1}\right\} $ in Fig. \ref{fig3} is VK-unstable in the
narrow area of the existence region, at $\mu $ close to $2$ in
Fig. \ref{fig4}(b). Other asymmetric solutions might be stable
according to the VK criterion.

For the antisymmetric solutions with $A=0$ and $B=-C\neq 0$, the
power curves are presented in Fig. \ref{fig4}(c). These solutions
also might be
stable according to the VK criterion in the whole existence region for $%
\varepsilon <1$. For higher values of $\varepsilon $, a
VK-unstable region
is located near the above-mentioned existence border, $\mu =2$ [see Eq. (\ref%
{eq8})].

\section{Numerical results}

The predictions of the VA were tested by solving stationary equations (\ref%
{eq3}), using an algorithm based on the modified Powell
minimization method \cite{nashi}. The initial guess for
constructing fundamental solitons centered at the interface of
three linked semi-infinite chains (Fig. \ref{fig1}) was taken in
the form of $u_{0}=v_{0}=w_{0}=A$ for symmetric modes
or $u_{0}=A,\,v_{0}=B,\,w_{0}=C$ for asymmetric ones (with different $A,B,C$%
), while the amplitudes of the lattice field at other sites were
set to be $0 $. Eventually, solutions different from symmetric
ones were found solely with $B=-C,\,A=0$, or with $A=C\neq B$, as
predicted analytically. The
results presented here are obtained for identical coupled chains of length $%
N=51$.

The stability of the stationary modes was first checked through
the linear-stability analysis. As a result, the eigenvalue (EV)
spectrum for modes of small perturbations was found, following the
procedure developed in Refs. \cite{symnash,nashi}. The
calculations were performed in parameter plane ($\varepsilon $,
$\mu $). These results were verified by direct numerical
simulations of the full system of equations (\ref{system}). The
simulations relied upon a numerical code which used the
sixth-order Runge-Kutta algorithm, as in Refs.
\cite{symnash,nashi}. The simulations were initialized by taking
stationary soliton profiles, to which random perturbations were
added.

\begin{figure}[h]
\center\includegraphics [width=6cm]{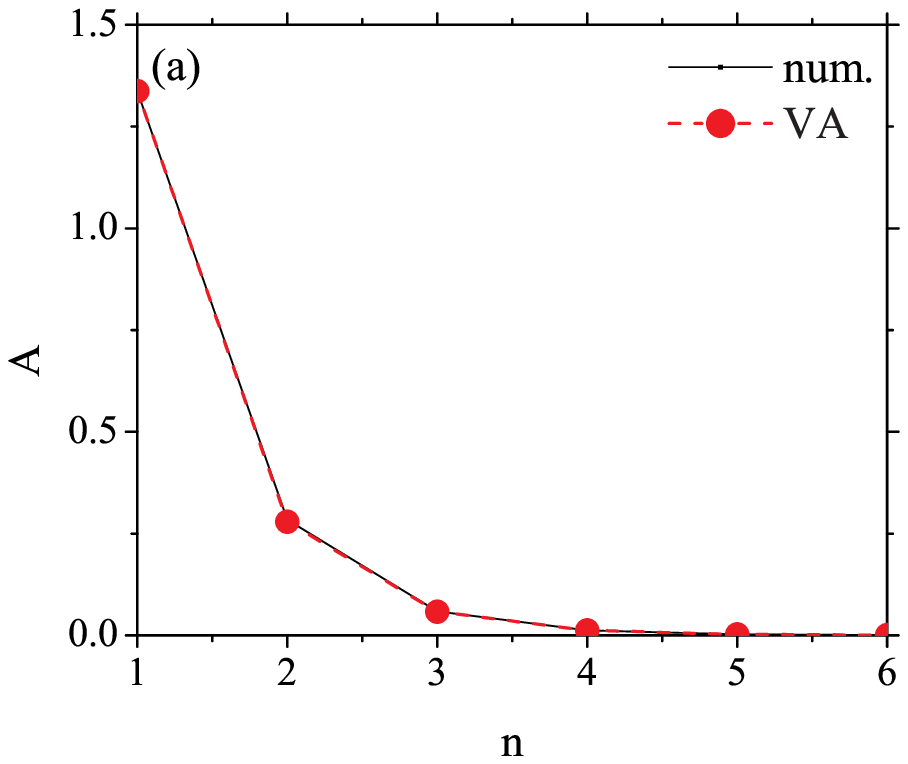}%
\includegraphics
[width=6cm]{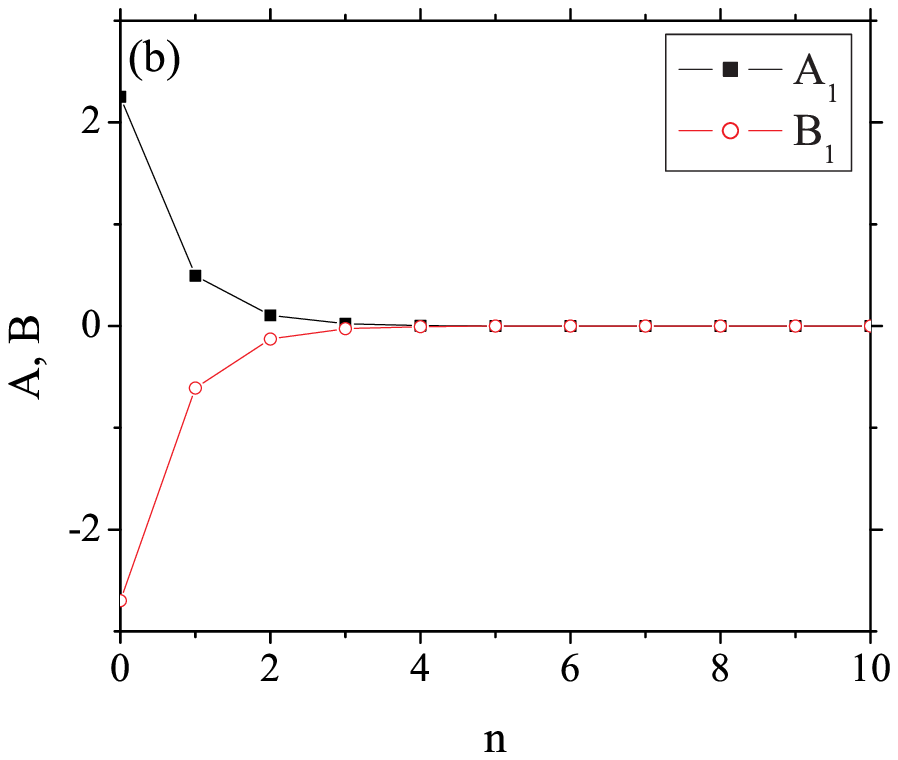} \includegraphics [width=6cm]{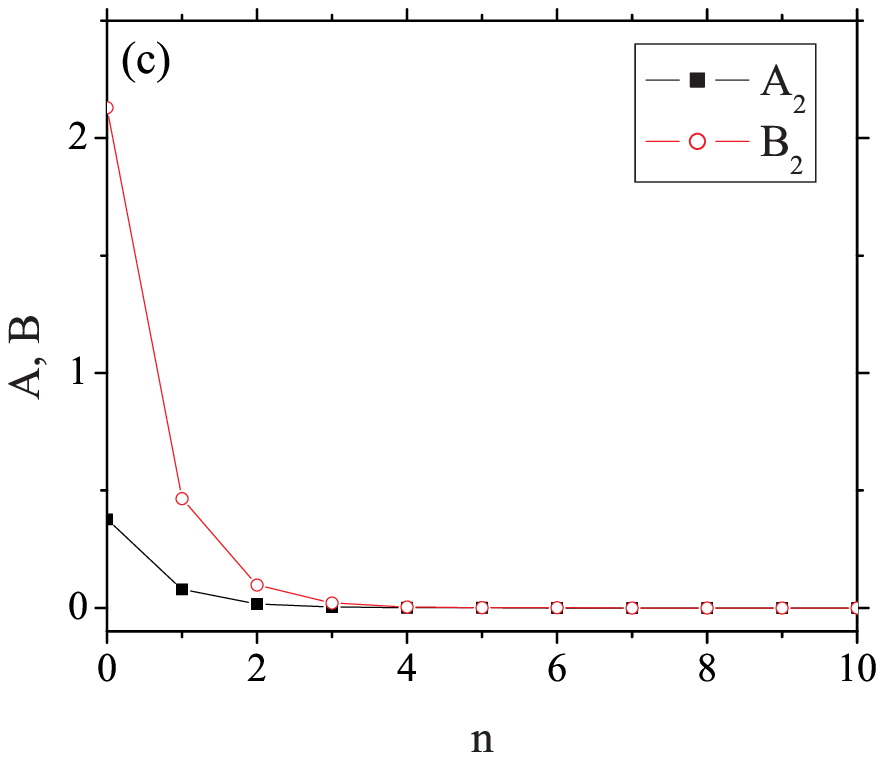}%
\includegraphics
[width=6cm]{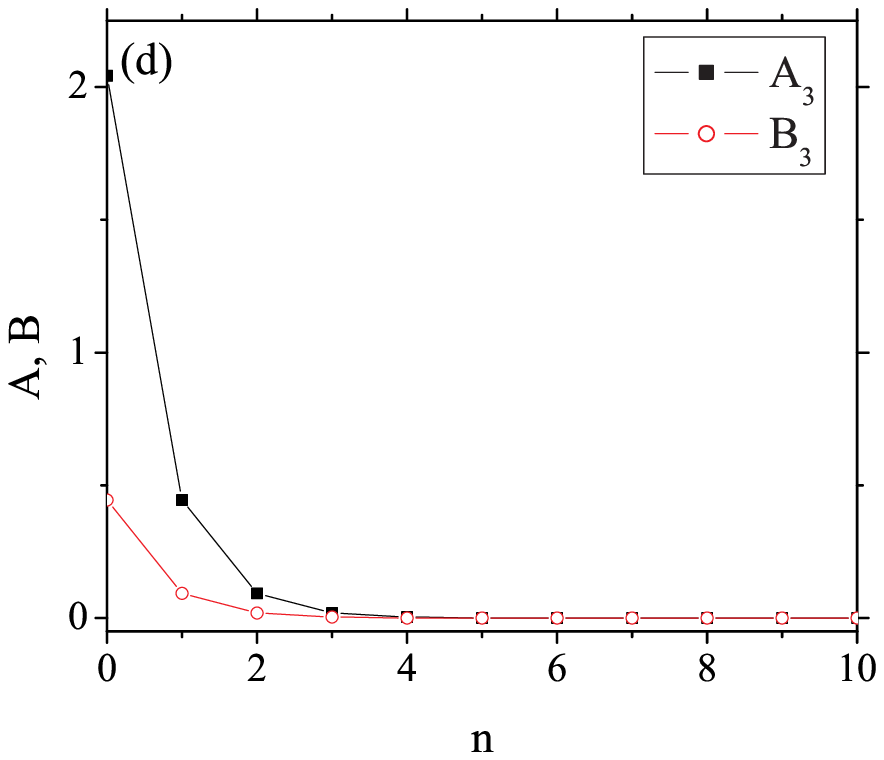} \caption{(Color online) Numerically
obtained profiles of the fundamental interface solitons: (a)
symmetric ($\protect\varepsilon =1.5$), (b)
asymmetric of type 1 ($\protect\varepsilon =1.5$), (c) asymmetric of type 2 (%
$\protect\varepsilon =0.7$), and (d) asymmetric of type 3 ($\protect%
\varepsilon =0.5$) (these types are defined in Fig.
\protect\ref{fig6}). Panel (a) depicts the numerically found SyS
profile (solid line with symbols) along with its numerically
predicted counterparts ("VA") (dashed line with symbols), while on
panels (b)-(d) the numerically found solution profiles are shown
by symbols and VA counterparts by solid lines.} \label{fig5}
  %fig4
\end{figure}

\begin{figure}[h]
\center\includegraphics [width=6cm]{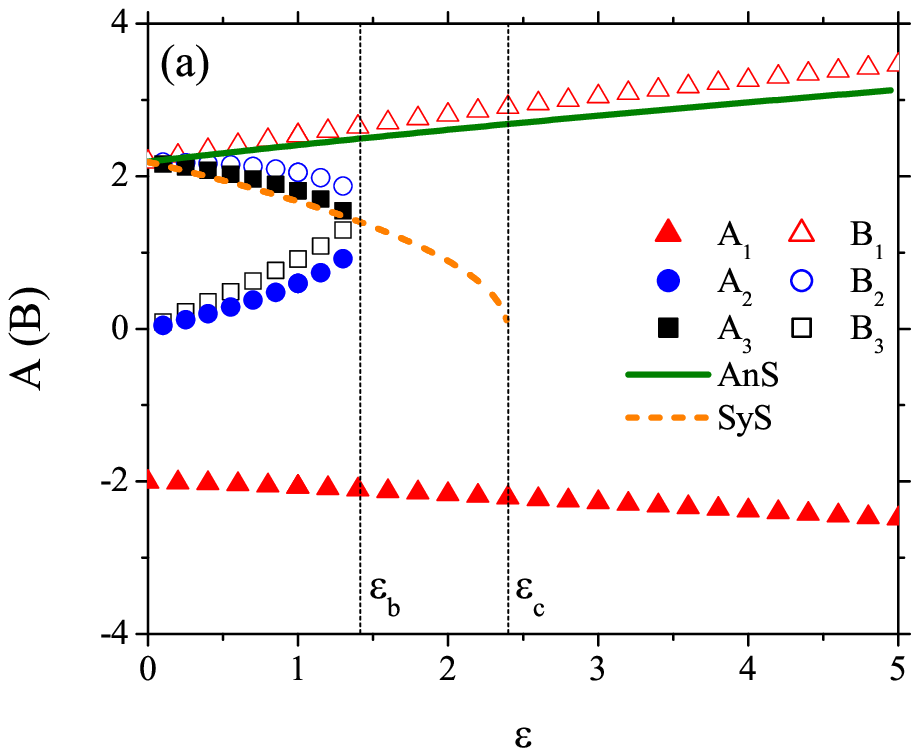}\includegraphics
[width=6cm]{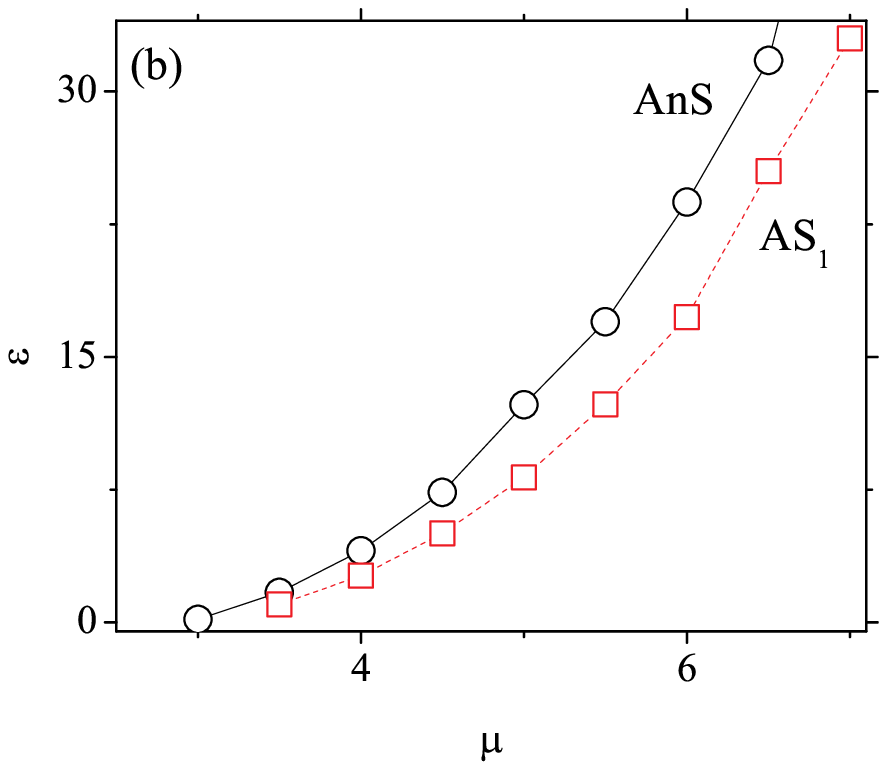} \caption{(Color online) (a) The amplitudes
of numerically generated symmetric solitons ("SyS", the orange
dashed line), and asymmetric ("AS") ones, corresponding to
branches (1), (2), and (3) -- red (triangles), blue (circles) and
black symbols (squares). These branches are counterparts of the
branches with the same numbers, which were predicted by the
variational approximation and displayed in Fig.
\protect\ref{fig3}. The solid olive line denotes the antisymmetric
branch with $A=0,B=-C\neq 0$, which also has its variational
counterpart in Fig. \protect\ref{fig3}. Another similarity to Fig.
\protect\ref{fig3} is that each solution has its negative mirror
image, only the solutions with positive amplitudes being
displayed here. The dotted lines denote the critical values $\protect%
\varepsilon _{\mathrm{b}}$ and $\protect\varepsilon _{\mathrm{c}}$
obtained
numerically. (b) The upper boundary of the existence region of the AnS and AS%
$_{1}$ complexes is plotted by the solid black line with circles
and red dashed line with squares, respectively. The corresponding
soliton complexes exist in the area below the boundary curves.}
\label{fig6}  %fig5
\end{figure}

\begin{figure}[h]
\center\includegraphics [width=7cm]{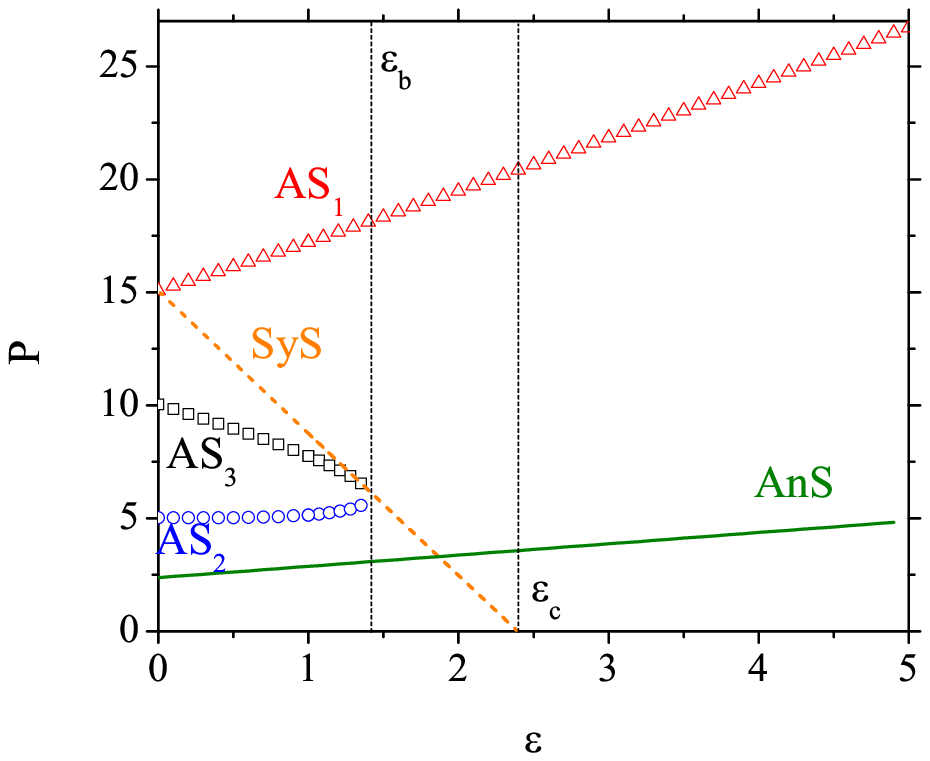} \caption{(Color
online) (a) The power of numerically generated symmetric solitons
("SyS", the orange dashed line), and asymmetric ("AS") ones,
corresponding to branches (1), (2), and (3) -- red (triangles),
blue (circles) and black symbols (squares). The solid olive line
denotes the antisymmetric branch
with $A=0,B=-C\neq 0$. The dotted lines denote the critical values $\protect%
\varepsilon _{\mathrm{b}}$ and $\protect\varepsilon _{\mathrm{c}}$
obtained numerically.}
\label{fig7}  %fig5
\end{figure}

Typical shapes of symmetric and asymmetric solitons found in the
numerical form are displayed in Fig. \ref{fig5}, well fitting the
corresponding VA predictions. The respective dependencies of the
solitons' amplitudes, A and B, on coupling constant $\varepsilon $
are displayed for all types of the solitons in Fig. \ref{fig6}
(a), while the corresponding dependencies $P(\varepsilon)$ are
shown in Fig. \ref{fig7}. From these figures it can be concluded
that both the amplitude and power characteristics give
qualitatively the same information about the localized surface
modes. The numerical results demonstrate that the symmetric and
three asymmetric branches coexist in certain bounded regions of
the parameter space, similar to what was predicted in Fig.
\ref{fig3}.

In general, the comparison of the numerical and variational
results demonstrates that the predictions of the VA for the
existence region of the symmetric and asymmetric isosceles
complexes of types $2$ and $3$ are very accurate. On the other
hand, the numerically found existence regions for the
antisymmetric solitons of type $3$ and antisymmetric complexes are
bounded, on the contrary to the prediction of the VA. For example,
the VA predicts
that the antisymmetric complexes can exist for arbitrary $\varepsilon $ and $%
\mu >2$, while the numerical calculation shown the appearance of
the upper limit with respect to $\varepsilon $ at fixed $\mu $,
see Fig. \ref{fig6} (b). The upper existence boundary for the
antisymmetric solutions was found at extremely high values of the
inter-chain coupling constant, $\varepsilon $ $\sim 100$.

\begin{figure}[h]
\center\includegraphics [width=4cm]{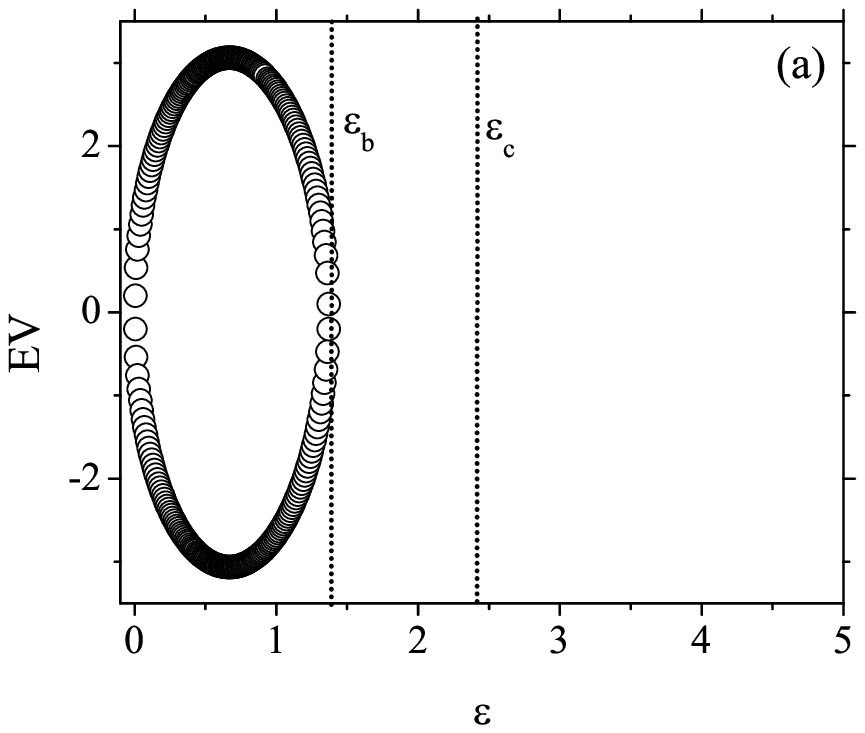}%
\includegraphics
[width=4cm]{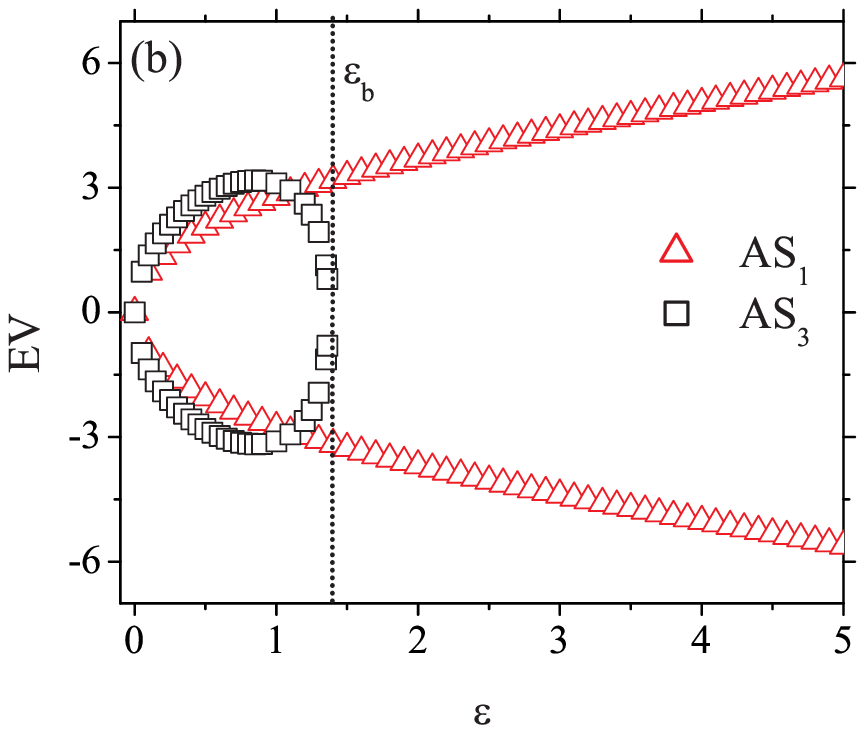}\includegraphics [width=4cm]{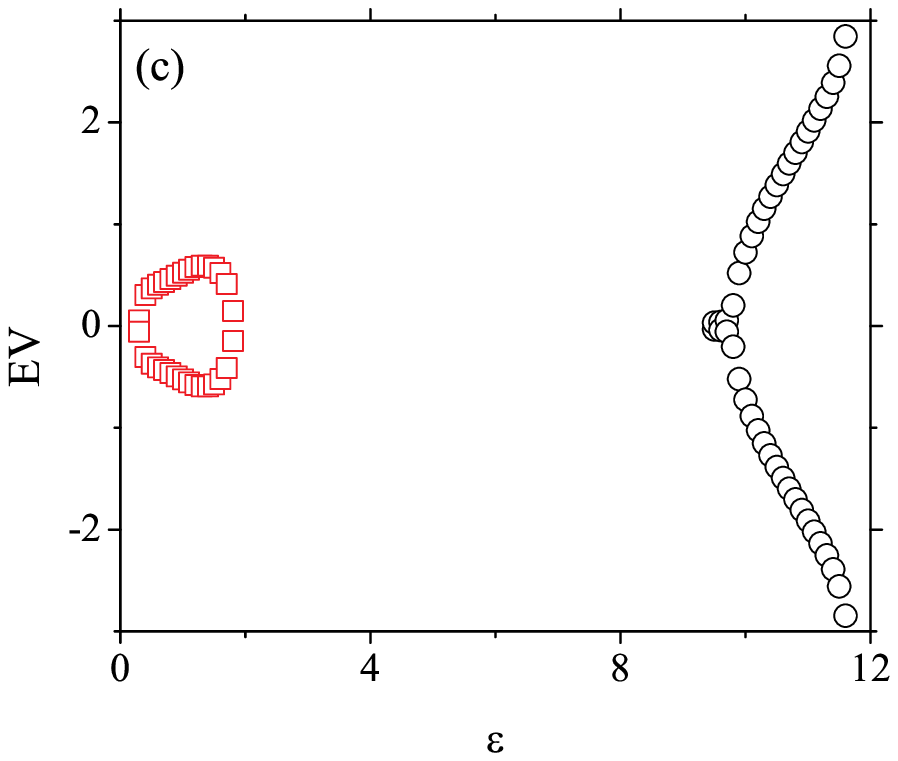}
\caption{(Color online) The purely real eigenvalues (EVs) versus
the interface coupling constant $\protect\varepsilon $ for fixed
$\protect\mu =5$ are shown in plots (a) and (b) for the symmetric
and the asymmetric ("AS") complexes of types $1$ and $3$,
respectively. The implication of the picture shown in panel (a) is
that the symmetric solitons
are \emph{stable} in the interval of $\protect\varepsilon _{\mathrm{b}}<%
\protect\varepsilon <\protect\varepsilon _{\mathrm{c}}$. In plot
(c), the real part of the complex EV (red squares) and pure real
EVs (black circles) are displayed for the antisymmetric solitons.
Another asymmetric complex (the one with amplitudes A$_{2}$,
B$_{2}$ in Figs. \protect\ref{fig3} and \protect\ref{fig6}) is
\emph{stable} in the entire existence region, $0\leq
\protect\varepsilon \leq \protect\varepsilon _{\mathrm{b}}$.}
\label{fig8}  %eigen
\end{figure}

The linear-stability analysis shows that a stability window exists
for the symmetric soliton complexes in the region between
$\varepsilon _{\mathrm{c}}$
[i.e., for $\mu \ $close to $\mu _{\mathrm{c}}=1/(2\varepsilon _{\mathrm{c}%
})+2\varepsilon _{\mathrm{c}}$, as it follows from Eq. (\ref{c})] and $%
\varepsilon _{\mathrm{b}}$, see Fig. \ref{fig8}(a). These
symmetric complexes are formed by solitons which are wider and
possess smaller amplitudes than their counterparts belonging to
unstable symmetric complexes.

\begin{figure}[h]
\center\includegraphics [width=4cm]{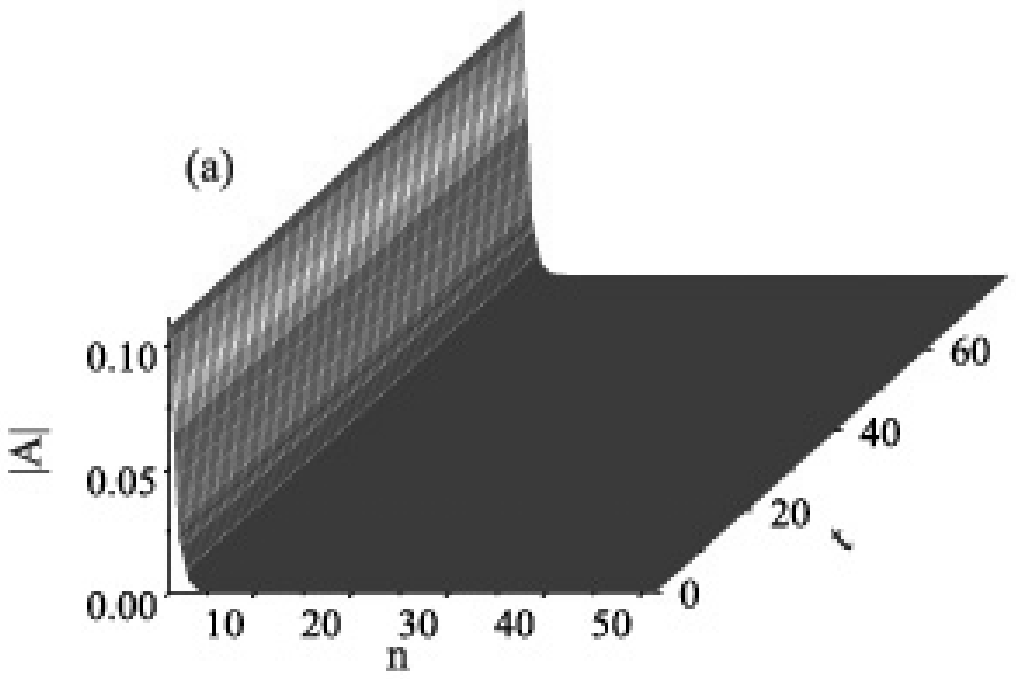}
\includegraphics
[width=4cm]{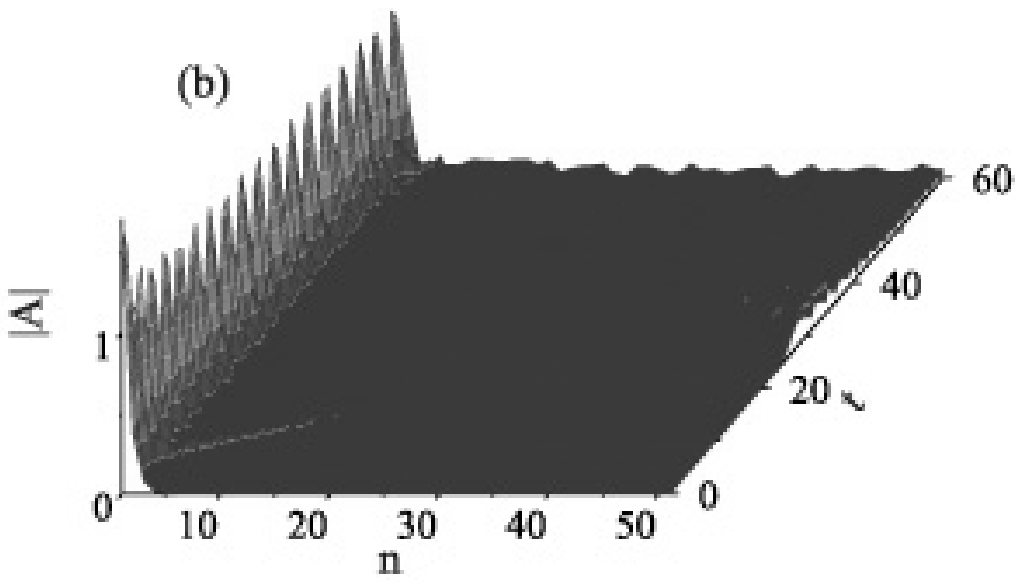}\includegraphics [width=4cm]{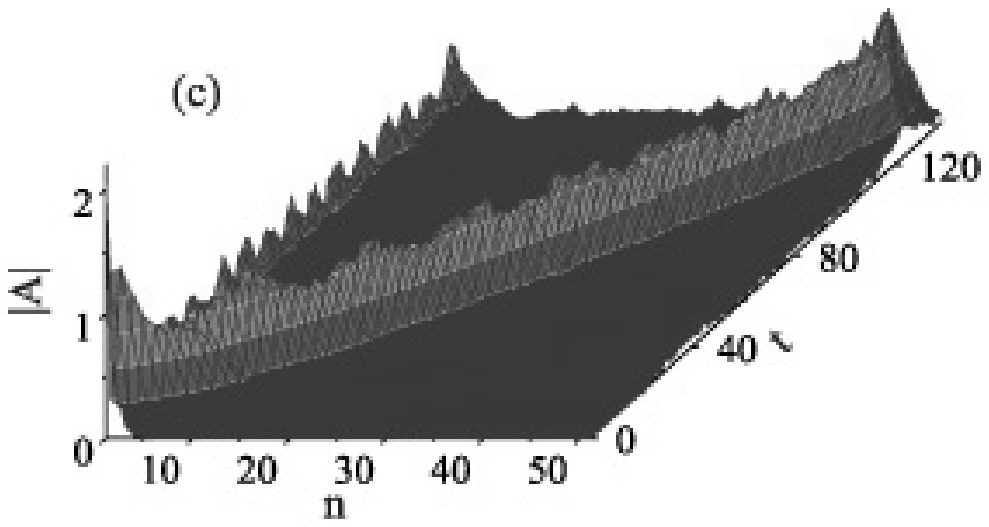}
\caption{Typical examples of the evolution of components of
perturbed three-soliton complexes: (a) a stable symmetric soliton with $%
\protect\varepsilon =1.69$; (b) an unstable symmetric complex with $\protect%
\varepsilon =1.09$; (c) a component of the unstable asymmetric
complex of type $1$ (see Fig. \protect\ref{fig3}) with
$\protect\varepsilon =1.69$. As well as in other figures, the
propagation constant of unperturbed solitons here is $\protect\mu
=5$.} \label{fig9}
\end{figure}

The dynamics of the stable symmetric soliton complex is illustrated by Fig. %
\ref{fig9}(a) which shows the evolution of its component. In the
rest of their existence region, the symmetric complexes are
unstable, their instability being accounted for by purely real EV
pairs. Under small perturbations, an unstable symmetric soliton
complex sheds off a part of its power and relaxes into a trapped
interface asymmetric breathing complex with a smaller power. The
dynamics of the component belonging to the unstable symmetric
complex is shown in Fig. \ref{fig9}(b).

The symmetric solitons with $\varepsilon =0$ correspond to the
usual surface solitons in uncoupled semi-infinite lattices. It is
well known that such surface modes are stable in almost the whole
existence region \cite{rad2}, which is provided by the balance
between the interaction of the soliton with the surface and the
bulk lattice. The instability of the symmetric soliton complex in
the coupled lattice system is related to a stronger repulsion from
the interface than the repulsion induced by the intra-lattice
potential energy barrier far from critical $\varepsilon
_{\mathrm{c}}$, thus enforcing the perturbed strongly pinned
soliton component solitons to shed off a part of their power. A
consequence is the formation of more stable interface asymmetric
breathing modes with a smaller power.

\begin{figure}[h]
\center\includegraphics [width=9cm]{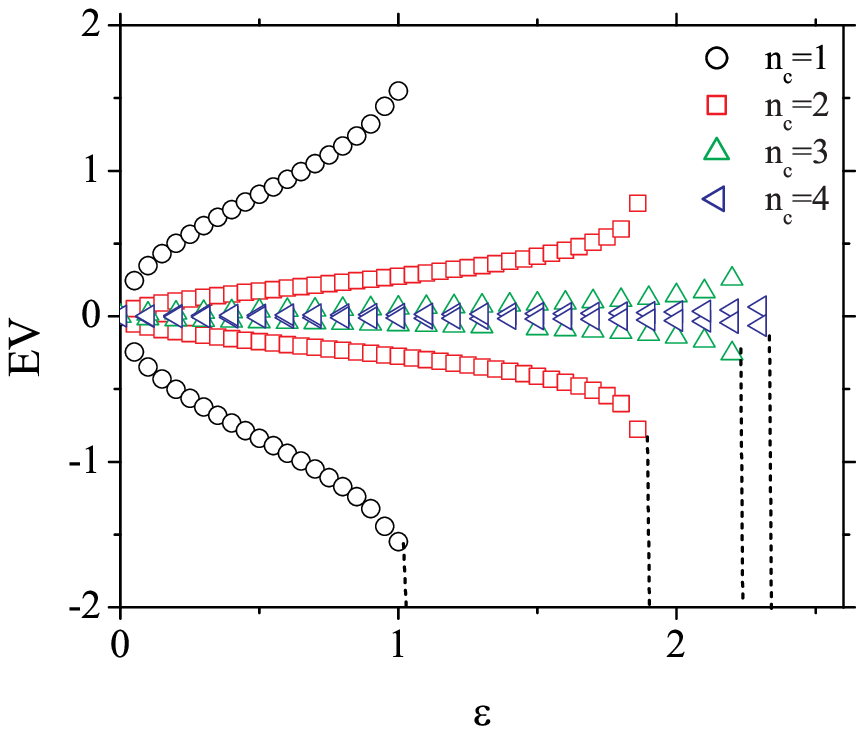} \caption{(Color
online) Purely real eigenvalues (EVs) versus the interface
coupling constant $\protect\varepsilon $ for fixed $\protect\mu
=5$. Fundamental solitons are centered at $n_{\mathrm{c}}=1,2,3,4$
in each chain. forming the symmetric configuration. Dotted lines
denote the boundary of the corresponding soliton-existence
regions. It is obvious that the solitons shifted deeper into the
lattices are more stable.} \label{fig10}
\end{figure}

The difference between the interface and intra-lattice potential
energies may also explain an enhanced stability of symmetric
complexes which are centered farther away from the interface. An
example is plotted in Fig. \ref{fig10}, where purely real EVs are
presented for symmetric soliton complexes whose components are
centered at distances $n_{\mathrm{c}}=1,2,3,4$ from the lattice
interface in the corresponding lattices. Eventually, the symmetric
complexes centered at $n_{c}>4$ are stable in their entire
existence region. We stress that only the soliton complexes of the
symmetric type were found, trying to set the centers of the
component solitons farther from the interface.

The EV spectra for the asymmetric isosceles solitons, which are
labeled by subscripts $1$ and $3$ in Figs. \ref{fig3} and
\ref{fig6}, indicate the exponential instability in their entire
existence regions, see Fig. \ref{fig8}(b). This conclusion is
confirmed by direct simulations, which show that perturbed
asymmetric complexes of these types radiate a significant part of
their power in the form of a breathing complex which moves across
the corresponding chains, see Fig. \ref{fig9}(c). The instability
of asymmetric solutions can be associated with the trend of the
system to relax toward an energetically preferable state in the
presence of the two above-mentioned forces -- the repulsion from
the lattice interface, and the force induced by the bulk lattice,
which is measured by the respective Peierls-Nabarro potential
barrier \cite{nashi}. The antisymmetric solutions with
$A=0,\,B=-C$ are shown to be unstable against oscillatory
perturbations in the region of $\varepsilon <\varepsilon _{b}$,
see Fig. \ref{fig8}(c), while a narrow region with exponentially
unstable antisymmetric complexes is found near the upper boundary
for their existence domain. Dynamical calculations confirm the
predictions of the linear stability analysis.

The isosceles soliton branch labeled by subscript $2$ in Figs.
\ref{fig3} and \ref{fig6}, which is \emph{stable} in the whole
existence region according to the linear-stability analysis, is
stable in direct simulations as well. Therefore, it can be
concluded that the symmetry-breaking bifurcation at $\varepsilon
=\varepsilon _{\mathrm{b}}$ is related to the stability exchange
between the destabilized symmetric complex and the two emerging
isosceles asymmetric complexes, one of which is stable and the
other exponentially unstable. In addition, the antisymmetric
complexes change their stability in the neighborhood of the
bifurcation point (from the oscillatory instability at
$\varepsilon <\varepsilon _{\mathrm{b}}$ to the \emph{stability}
at $\varepsilon >\varepsilon _{\mathrm{b}}$).

Returning to the global existence diagrams, it is worthy to note
that two areas with coexisting symmetric, asymmetric and
antisymmetric solutions can be identified: the domain featuring
the coexistence of stable symmetric, unstable asymmetric and
stable antisymmetric complexes, and the domain where the
symmetric, antisymmetric and two asymmetric complexes are unstable
and one isosceles solution branch is stable.

As said above, the dynamics of the interface soliton complexes is
strongly related to the balance of the interface and bulk-lattice
potential energies. When the inter-lattice coupling is too small,
the interface complexes of both the symmetric and asymmetric types
formed by the fundamental solitons are unstable. The exception is
(as actually mentioned above) the isosceles
asymmetric complex with $B>A=C$ (mode $2$ in Fig. \ref{fig3} and \ref{fig6}%
), which is stable in its entire existence region. The strong
repulsion from the interface, in comparison to the force from the
bulk lattice, makes the formation of stable surface solitons
difficult. By increasing the inter-lattice coupling, the energy
difference between the interface and intra-lattice forces gets
smaller, which makes it possible to create stable localized
interface symmetric and antisymmetric complexes. With the further
increase of the inter-lattice linkage against the intra-lattice
coupling, which means proceeding to $\varepsilon >2$ for fixed
$\mu $, the highly unstable asymmetric complex and stable
antisymmetric one are the only soliton species generated by the
system. These findings are correlated with results reported for
the two-chain version of the present system in Ref.
\cite{symnash}, where the stable symmetric, antisymmetric, and
antisymmetric branches have been found in a certain part of the
corresponding existence regions.

\section{Conclusion}

In this paper, we have analyzed the properties of fundamental
localized interface complexes excited at the interconnection of
three nonlinear semi-infinite chains. This system may be realized
in nonlinear optics and BEC. In the framework of the system of
three semi-infinite DNLS equations with the self-focusing on-site
nonlinearity, coupled at the single site, we have found, by means
of the VA (variational approximation) and numerical calculations,
the threshold value of the inter-lattice coupling constant
bounding the creation of symmetric surface soliton complexes. The
existence of families of symmetric and asymmetric interface
soliton complexes has been demonstrated. The variational
predictions for the stationary modes were checked numerically. The
stability analysis shows that the symmetric soliton complexes,
which are created as a stable solution branch at the critical
value of the lattice inter-coupling parameter, destabilize at the
bifurcation point, where two isosceles asymmetric soliton
complexes are created, one of them stable and the other
exponentially unstable. In addition, the stability of the
antisymmetric complexes changes twice -- at the bifurcation point
and near the upper boundary of their existence region. The third
isosceles solution branch is exponentially unstable in its entire
existence region. In other words, in the trilete system, the
symmetric solution branch, three asymmetric branches and the
antisymmetric one coexist in a part of the parameter space (past
the symmetry-breaking bifurcation). These solution branches exist
with their mirror-image counterparts. Direct simulations
demonstrate that unstable symmetric complexes are transformed,
radiating away a part of their power, into robust oscillating
modes in the form of localized interface breathing complexes,
while the unstable asymmetric complexes form breathers traveling
away from the interface through the lattice. The origin of this
behavior of the surface modes is related to the balance between
the repulsion from the surface and the intra-lattice interactions.

Because the instability of asymmetric soliton complexes is
stipulated by the repulsion from the interface (the linkage site,
$n=0$), it seems plausible that they may be stabilized by making
the on-site nonlinearity stronger at this site. Another
interesting possibility is to consider solitons of the vortex type
(cf. vortices studied in other linearly-coupled tri-core systems
\cite{20,Skryabin}). The respective ansatz may be taken as
$\left\{
u_{n},v_{n},w_{n}\right\} =A\left\{ 1,e^{i\pi /3},e^{2i\pi /3}\right\} \exp {%
(-an)}$, cf. ansatz (\ref{eq6}). This vortex may be classified as
one of the
"off-site" type, as its virtual pivot is located between the sites.

\section*{Acknowledgments} Lj.H. and A.M. acknowledge support from the
Ministry of Education and Science, Serbia (Project III45010).


%% References


\bibliographystyle{elsarticle-num}
\bibliography{<your-bib-database>}



\end{document}